\documentclass[sigconf]{acmart}

\usepackage{acmart-taps}
\usepackage{multirow}
\usepackage{colortbl}
\usepackage{makecell}
\usepackage{dblfloatfix}
\usepackage{balance}

\definecolor{green}{rgb}{0.1,0.1,0.1}
\newcommand{\mycomment}[1]{}

\AtBeginDocument{%
  \providecommand\BibTeX{{%
    \normalfont B\kern-0.5em{\scshape i\kern-0.25em b}\kern-0.8em\TeX}}}

\setcopyright{acmlicensed}
\copyrightyear{2023}
\acmYear{2023}
\acmDOI{10.1145/3586183.3606803}
\acmConference[UIST '23]{The 36th Annual ACM Symposium on User Interface Software and Technology}{October 29-November 1, 2023}{San Francisco, CA, USA}
\acmBooktitle{The 36th Annual ACM Symposium on User Interface Software and Technology (UIST '23), October 29-November 1, 2023, San Francisco, CA, USA}
\acmPrice{15.00}
\acmDOI{10.1145/3586183.3606803}
\acmISBN{979-8-4007-0132-0/23/10}

\begin{document}


\title{STAR: Smartphone-analogous Typing in Augmented Reality}


\author{Taejun Kim}
\authornote{The author is also affiliated with HCI Lab, KAIST, Daejeon, Republic of Korea. This work was done while the author was a Research Intern at Meta.}
\affiliation{%
  \institution{Realiy Labs Research, Meta}
  \city{Toronto, ON}
  \country{Canada}}
\email{taejun.kim@kaist.ac.kr}

\author{Amy Karlson}
\affiliation{%
  \institution{Reality Labs Research, Meta}
  \city{Redmond, WA}
  \country{USA}
}
\email{akkarlson@meta.com}

\author{Aakar Gupta}
\affiliation{%
  \institution{Reality Labs Research, Meta}
  \city{Redmond, WA}
  \country{USA}
}
\email{aakar.hci@gmail.com}

\author{Tovi Grossman}
\affiliation{%
  \institution{University of Toronto}
  \city{Toronto, ON}
  \country{Canada}
}
\email{tovi@dgp.toronto.edu}
\author{Jason Wu}
\authornote{The author is also affiliated with Carnegie Mellon University, Pittsburgh, PA, USA. This work was done while the author was a Research Intern at Meta.}
\affiliation{%
  \institution{Reality Labs Research, Meta}
  \city{Toronto, ON}
  \country{Canada}}
\email{jsonwu@cmu.edu}

\author{Parastoo Abtahi}
\affiliation{%
  \institution{Reality Labs Research, Meta}
  \city{Toronto, ON}
  \country{Canada}
}
\email{parastoo@princeton.edu}

\author{Christopher Collins}
\affiliation{%
  \institution{Reality Labs Research, Meta}
  \city{Toronto, ON}
  \country{Canada}
}
\email{chriscollins@meta.com}

\author{Michael Glueck}
\affiliation{%
  \institution{Reality Labs Research, Meta}
  \city{Toronto, ON}
  \country{Canada}
}
\email{mglueck@meta.com}

\author{Hemant Bhaskar Surale}
\affiliation{%
  \institution{Reality Labs Research, Meta}
  \city{Toronto, ON}
  \country{Canada}
}
\email{hemantsurale@meta.com}


\renewcommand{\shortauthors}{Kim et al.}

\begin{abstract}
While text entry is an essential and frequent task in Augmented Reality (AR) applications, devising an efficient and easy-to-use text entry method for AR remains an open challenge. This research presents STAR, a smartphone-analogous AR text entry technique that leverages a user's familiarity with smartphone two-thumb typing. With STAR, a user performs thumb typing on a virtual QWERTY keyboard that is overlain on the skin of their hands. During an evaluation study of STAR, participants achieved a mean typing speed of 21.9 WPM (i.e., 56\% of their smartphone typing speed), and a mean error rate of 0.3\% after 30 minutes of practice. We further analyze the major factors implicated in the performance gap between STAR and smartphone typing, and discuss ways this gap could be narrowed.
 \end{abstract}




\begin{teaserfigure}
  \centering
  \includegraphics[width=0.985\textwidth]{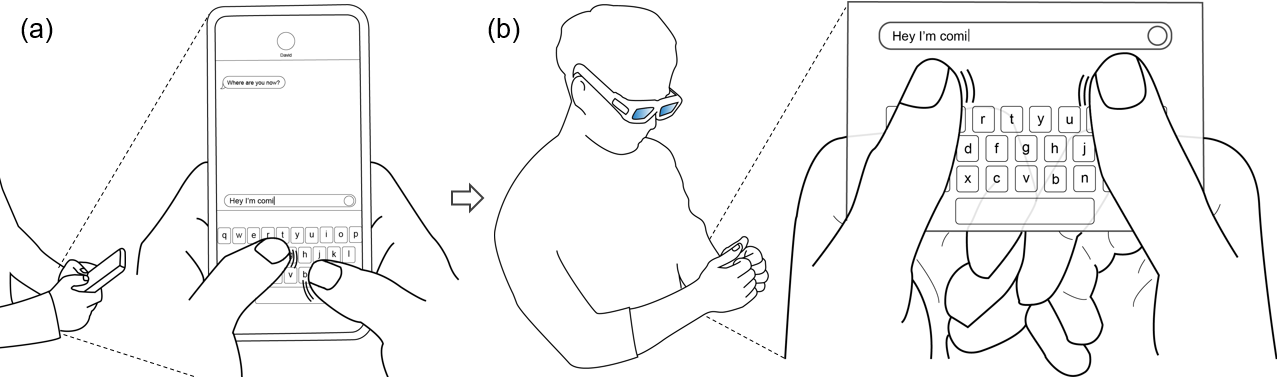}
  \caption{(a) Traditional physical smartphone typing and (b) STAR, a bare-hand, two-thumb text entry method in augmented reality. STAR transfers the two-thumb typing skills learned from using a physical smartphone to bare-hand AR typing.}
  \label{fig:teaser}
\end{teaserfigure}

\maketitle

\section{Introduction}

Recent advancements in Augmented Reality (AR) hardware and display technologies (e.g., glasses~\cite{googleglass, microsoftHololens, snapSpectacles}) have opened up new opportunities for utilizing AR in various fields such as remote work, health care, and education~\cite{roltgen2020classification, surale2022arcall, bipat2019analyzing}. However, despite the growing need for text entry in AR environments, devising an efficient and easy-to-use text entry solution remains an open challenge.

While researchers have developed several techniques to enable bare-hand text entry for unrestrained AR activities, prior bare-hand AR text entry techniques such as mid-air hand typing~\cite{markussen2014vulture, yi2015atk, dudley2018fast}, eye typing~\cite{kurauchi2016eyeswipe, lu2021itext}, or tapping on a fingertip~\cite{xu2019tiptext, xu2020bitiptext} are unfamiliar to users as they are drastically different from today's typing methods. In addition, common bare-hand AR text entry techniques often lack the haptic feedback delivered when one touches a physical surface. For example, the Microsoft Hololens 2 AR text entry method~\cite{dudley2018fast} requires users to touch a mid-air virtual keyboard using their fingers, without any haptic feedback. This lack of surface haptics has been found to impact input accuracy, typing speeds, and user fatigue~\cite{cheng2022comfortable, arora2017experimental, jang2017modeling, hincapie2014consumed}.

Herein, we present STAR, a novel bare-hand AR text entry technique that capitalizes on users' familiarity with two-thumb typing (i.e., both hands holding a smartphone while both thumbs are typing). Due to the ubiquity of smartphones, two-thumb typing is a familiar and rapid form of input for most people~\cite{palin2019people, smith2015us, wang2016smartphone}, even approaching the performance of desktop QWERTY keyboard typing for some users~\cite{feit2016we, palin2019people}. Our work explores how this familiar skill could be transferred to a bare-hand AR text entry. While it may not be possible to match the typing performance of smartphones with a bare-hand technique, we believe the performance gap between the two can be closed by leveraging the same skills of two-thumb typing.

To use STAR, one forms a ``knuckle posture''  with their hands (Figure~\ref{fig:teaser}b) as if holding a smartphone, which triggers the display of a mini virtual QWERTY keyboard on the user's hand through the Head-Mounted Display (HMD) they are wearing. STAR then enables the user to leverage the haptic feedback of touching their own skin to two-thumb type on this keyboard. If the user wishes to transition onto other tasks, they can seamlessly release the knuckle posture and the keyboard vanishes. The knuckle posture is an explicit mode-switching technique that prevents false activations. Further, STAR is expected to be more socially acceptable compared to other bare-hand text entry techniques because the smartphone-typing gesture naturally conveys the user's state of typing to others~\cite{ens2015candid, xia2022iteratively}. 

In an empirical evaluation, STAR showed an efficient text entry performance with a mean typing speed of 21.9 words per minute (WPM) and a mean error rate of 0.3\%. This is 56\% of smartphones typing speed, after only 30 minutes of practice. We further discuss ways to narrow the performance gap between STAR and physical smartphone typing based on the collected typing data and subjective user feedback.

The main contributions are as follows:

\begin{itemize}
  \item We present STAR, a novel on-skin, smartphone-analogous AR text entry technique that leverages existing typing skills from a physical smartphone.
  \item We explore the transferability of two-thumb typing skills to AR. In an elicitation study with 29 participants, three different hand postures were discovered for typing on an imaginary smartphone on their skin. These postures, along with other design parameters, were then tested to ensure sufficient skill transfer from smartphone typing to STAR.
  \item An analysis of the performance delta between STAR and smartphone typing reveals that improved hand tracking and thumb tap sensing may provide an opportunity for STAR to achieve typing performance closer to that of a physical smartphone.
\end{itemize}

\aptLtoX{\begin{table*}[t]
  \setlength{\tabcolsep}{4pt} 
  \begin{tabular}{p{3.1cm}ccccccc}
    \toprule
    \rowcolor{gray!30}& & \multicolumn{2}{c}{\cellcolor{gray!30}\textbf{Typing Level}} & & & & \\    
    \rowcolor{gray!30} 
\multirow{-2}{*}{\textbf{Technique}} & \multirow{-2}{*}{\textbf{Modality}} & letter & word & \multirow{-2}{*}{\makecell{\textbf{Speed} \\ \textbf{(WPM)}}} & \multirow{-2}{*}{\makecell{\textbf{Error} \\ \textbf{(\%)}}} & \multirow{-2}{*}{\textbf{Training Amount}} & \multirow{-2}{*}{\textbf{Traits}} \\
    \midrule
    Controller Pointing \cite{speicher2018selection}  & \multirow{2}{* }{\makecell{Handheld controllers}} & \checkmark & & 15.4 & 1.0 & 5 minutes & \multirow{2}{*}{\makecell{Need to  hold controllers }} \\
    Word-Gesture \cite{chen2019exploring} & & & \checkmark & 16.4 & 15.6\textsuperscript{C} & None & \\
    \hline
     EyeSwipe \cite{kurauchi2016eyeswipe} & \multirow{3}{*}{Head + Eye} & \checkmark & \checkmark & 11.7 & 1.3 & 30 minutes & \multirow{3}{*}{\makecell{Impacts natural viewing,  induce eye fatigue}} \\
    iText \cite{lu2021itext} & & & \checkmark & 13.8 & 1.5\textsuperscript{W} & 72 phrases (4 days) & \\
    GlanceWriter \cite{cui2023glancewriter} & & \checkmark & \checkmark & 10.9 & 2.7\textsuperscript{W} & 10 minutes &   \\
    \hline
    Speech Recognition \cite{ruan2018comparing} & Speech & & \checkmark & 179 & 4.4 & 10 phrases & \makecell{Raise privacy concern}  \\    
    \hline
    Vulture \cite{markussen2014vulture} & \multirow{5}{*}{\makecell{Mid-air hand  movement}} & & \checkmark & 28.1 & 1.7\textsuperscript{W} & 48 phrases & \multirow{5}{*}{\makecell{Induce arm fatigue,  lack haptic feedback}}  \\
    ATK \cite{yi2015atk} & & & \checkmark & 29.2 & 0.4\textsuperscript{W} & 5 minutes + 45 phrases &  \\
    FastType \cite{sridhar2015investigating} & & \checkmark &  & 22.3\textsuperscript{P} & 2.3 & 15 minutes &  \\
    VISAR \cite{dudley2018fast} & & \checkmark & \checkmark & 17.8 & 0.6 & 80 phrases &  \\   
     ThumbAir \cite{gil2023thumbair}  & & & \checkmark & 13.7 & 1.2 & 140 words + 35 phrases &  \\
    \hline
    QwertyRing \cite{gu2020qwertyring}  & \multirow{3}{* }{\makecell{Finger taps on  physical surface}} & & \checkmark & 20.6 & 1.3 & 120 phrases (4 days) & \multirow{3}{*}{\makecell{Rely on  probabilistic decoder }} \\
    TapType \cite{streli2022taptype} & & \checkmark & \checkmark & 19.2/9.0\textsuperscript{L} & 0.6/0.4\textsuperscript{L} & 35 phrases & \\
    TypeAnywhere \cite{streli2022taptype} & & & \checkmark & 70.6 & 1.5 & 120 m. + 80 ph. (4 days) & \\
    \hline    
    PalmType \cite{wang2015palmtype} &  \multirow{4}{*}{\makecell{On-skin  finger touch}} & \checkmark & & 7.7 & 1.6 & (not specified) & \multirow{4}{*}{\makecell{Require users to learn  new interaction paradigms,  use custom hardware}}  \\
    DigiTouch \cite{whitmire2017digitouch} & & \checkmark & \checkmark & 16.0 & 0.9 & 180 minutes & \\
    TipText \cite{xu2019tiptext} & & & \checkmark & 13.3 & 0.3 & 30 phrases &  \\
    BiTipText \cite{xu2020bitiptext} & & & \checkmark & 25 & 0.03 & 30 phrases &   \\    
    \hline
    \textbf{STAR} & \textbf{\makecell{On-skin  thumb taps}} & \checkmark & \checkmark & \textbf{21.9} & \textbf{0.3} & \textbf{30 minutes} & \textbf{\makecell{Leverages familiar  smartphone typing skills}}\\ 
    \bottomrule
  \end{tabular}
  \caption{Overview of text entry techniques proposed for, or potentially usable in, HMD-based AR. The table compares these techniques along the following dimensions: modality, supported typing level (i.e., letter-level or word-level), mean typing speed, mean error rate, amount of training needed, and traits of the technique. Error indicates character-level uncorrected error rates unless specified (C denotes character-level corrected error rates, and W denotes word-level uncorrected error rates). P denotes peak typing speed that was measured by a repeated typing of the same word. L denotes mean typing speed with test phrases including Out-Of-Vocabulary (OOV) words, requiring letter-by-letter typing. It is important to note that the performance difference between each method may also be influenced by additional factors such as the input technology (i.e., sensing accuracy), the specific set of test phrases, the probabilistic decoder implemented, or the keyboard layout used. }
    \label{table:relatedWorkTable}
\end{table*}}{\begin{table*}[t]
  \setlength{\tabcolsep}{4pt} 
  \begin{tabular} {p{3.1cm}ccccccc}
    \hline
    \rowcolor{gray!30} & & \multicolumn{2}{c}{\textbf{Typing Level}} & & & & \\    
    \rowcolor{gray!30} \multicolumn{1}{c}{\multirow{-2}{*}{\textbf{Technique}}} & \multirow{-2}{*}{\textbf{Modality}} & letter & word & \multirow{-2}{*}{\makecell{\textbf{Speed} \\ \textbf{(WPM)}}} & \multirow{-2}{*}{\makecell{\textbf{Error} \\ \textbf{(\%)}}} & \multirow{-2}{*}{\textbf{Training Amount}} & \multirow{-2}{*}{\textbf{Traits}} \\
    \hline
    Controller Pointing \cite{speicher2018selection}  & \multirow{2}{* }{\makecell{Handheld \\ controllers}} & \checkmark & & 15.4 & 1.0 & 5 minutes & \multirow{2}{*}{\makecell{Need to \\ hold controllers }} \\
    Word-Gesture \cite{chen2019exploring} & & & \checkmark & 16.4 & 15.6\textsuperscript{C} & None & \\
    \midrule
     EyeSwipe \cite{kurauchi2016eyeswipe} & \multirow{3}{*}{Head + Eye} & \checkmark & \checkmark & 11.7 & 1.3 & 30 minutes & \multirow{3}{*}{\makecell{Impacts natural viewing, \\ induce eye fatigue}} \\
    iText \cite{lu2021itext} & & & \checkmark & 13.8 & 1.5\textsuperscript{W} & 72 phrases (4 days) & \\
    GlanceWriter \cite{cui2023glancewriter} & & \checkmark & \checkmark & 10.9 & 2.7\textsuperscript{W} & 10 minutes &   \\
    \midrule
    Speech Recognition \cite{ruan2018comparing} & Speech & & \checkmark & 179 & 4.4 & 10 phrases & \makecell{Raise privacy concern}  \\    
    \midrule
    Vulture \cite{markussen2014vulture} & \multirow{5}{*}{\makecell{Mid-air hand \\ movement}} & & \checkmark & 28.1 & 1.7\textsuperscript{W} & 48 phrases & \multirow{5}{*}{\makecell{Induce arm fatigue, \\ lack haptic feedback}}  \\
    ATK \cite{yi2015atk} & & & \checkmark & 29.2 & 0.4\textsuperscript{W} & 5 minutes + 45 phrases &  \\
    FastType \cite{sridhar2015investigating} & & \checkmark &  & 22.3\textsuperscript{P} & 2.3 & 15 minutes &  \\
    VISAR \cite{dudley2018fast} & & \checkmark & \checkmark & 17.8 & 0.6 & 80 phrases &  \\   
     ThumbAir \cite{gil2023thumbair}  & & & \checkmark & 13.7 & 1.2 & 140 words + 35 phrases &  \\
    \midrule
    QwertyRing \cite{gu2020qwertyring}  & \multirow{3}{* }{\makecell{Finger taps on \\ physical surface}} & & \checkmark & 20.6 & 1.3 & 120 phrases (4 days) & \multirow{3}{*}{\makecell{Rely on \\ probabilistic decoder }} \\
    TapType \cite{streli2022taptype} & & \checkmark & \checkmark & 19.2/9.0\textsuperscript{L} & 0.6/0.4\textsuperscript{L} & 35 phrases & \\
    TypeAnywhere \cite{streli2022taptype} & & & \checkmark & 70.6 & 1.5 & 120 m. + 80 ph. (4 days) & \\
    \midrule    
    PalmType \cite{wang2015palmtype} &  \multirow{4}{*}{\makecell{On-skin \\ finger touch}} & \checkmark & & 7.7 & 1.6 & (not specified) & \multirow{4}{*}{\makecell{Require users to learn \\ new interaction paradigms, \\ use custom hardware}}  \\
    DigiTouch \cite{whitmire2017digitouch} & & \checkmark & \checkmark & 16.0 & 0.9 & 180 minutes & \\
    TipText \cite{xu2019tiptext} & & & \checkmark & 13.3 & 0.3 & 30 phrases &  \\
    BiTipText \cite{xu2020bitiptext} & & & \checkmark & 25 & 0.03 & 30 phrases &   \\    
    \midrule
    \textbf{STAR} & \textbf{\makecell{On-skin \\ thumb taps}} & \checkmark & \checkmark & \textbf{21.9} & \textbf{0.3} & \textbf{30 minutes} & \textbf{\makecell{Leverages familiar \\ smartphone typing skills}}\\ 
    \bottomrule
  \end{tabular}
  \caption{Overview of text entry techniques proposed for, or potentially usable in, HMD-based AR. The table compares these techniques along the following dimensions: modality, supported typing level (i.e., letter-level or word-level), mean typing speed, mean error rate, amount of training needed, and traits of the technique. Error indicates character-level uncorrected error rates unless specified (C denotes character-level corrected error rates, and W denotes word-level uncorrected error rates). P denotes peak typing speed that was measured by a repeated typing of the same word. L denotes mean typing speed with test phrases including Out-Of-Vocabulary (OOV) words, requiring letter-by-letter typing. It is important to note that the performance difference between each method may also be influenced by additional factors such as the input technology (i.e., sensing accuracy), the specific set of test phrases, the probabilistic decoder implemented, or the keyboard layout used. }
    \label{table:relatedWorkTable}
\end{table*}}

\section{Related Work}

We first review the literature on large-scale datasets that have been used to characterize smartphone typing skills. Subsequently, we discuss on-body interactions and skill transfer. Lastly, we scrutinize text entry techniques proposed for HMD-based AR and position STAR among them. 

\subsection{Smartphone Typing Skills}

Due to the ubiquity of smartphones, smartphone typing has become a familiar and rapid form of input for most users~\cite{palin2019people, smith2015us, wang2016smartphone}. Common smartphone text entry techniques involve the coordinated tapping of user's thumbs on a miniature QWERTY keyboard.

Palin et al.~\cite{palin2019people} presented a large-scale dataset of mobile text entry input collected from a web-based transcription task with 37,370 volunteers (mean age: 24.1). The mean typing speed was 36.2 WPM with 2.3\% uncorrected errors and the fastest typists reached over 80 WPM, approaching the performance levels of desktop QWERTY keyboard typing, which has been found to have a mean typing speed of 51.6 WPM~\cite{dhakal2018observations} and the fastest speed up to 130 WPM~\cite{pinet2022typing}. Over 82\% of participants used two thumbs to type, which was significantly faster than using one finger (i.e., 37.7 vs. 29.2 WPM). The mean typing speed was similar between genders (i.e., \textasciitilde36.1 WPM for both men and women) but differed between age groups (i.e., 39.6 WPM for 10-19 years old vs. 26.3 WPM for 50-59 years old). Although it may not be possible to match the same typing performance with bare-hand techniques, STAR aims to narrow the gap by leveraging the pervasive skills of smartphone two-thumb typing.

\subsection{On-Body Interactions and Skill Transfer}

Using the body surface as an input space has been a long-standing interaction paradigm. For example, Harrison et al.~\cite{harrison2010skinput} presented Skinput, an interaction using skin as an always-available input surface. Imaginary Phone~\cite{gustafson2011imaginary} proposed using the skin on one's hand as a touch input surface and suggested ``transfer learning'', i.e., by using a physical smartphone, a user inadvertently learns the interface and can then transfer that knowledge to an imaginary interface on the surface of their hand. Gustafson et al.~\cite{gustafson2011imaginary} noted that \textit{``the transfer model is viable, even though full accuracy will not be redeemed until higher resolution tracking equipment becomes available''}. STAR leverages this same concept by supporting the transfer of learning from physical smartphone typing to virtual bare-hand typing.

\subsection{Text Entry in HMD-based AR}

The proposed text entry techniques for HMDs were often motivated by AR and VR applications, with some techniques~\cite{speicher2018selection, kurauchi2016eyeswipe, dudley2018fast} being usable in both domains. Herein, we review text entry techniques proposed for, or potentially usable in, HMD-based AR systems which leverage input modalities such as controller, gaze, speech, mid-air hand movements, finger tapping on physical surfaces, or on-skin touches (Table 1).

\subsubsection{Handheld Controllers}

Text entry techniques using handheld controllers~\cite{speicher2018selection, boletsis2019controller, chen2019exploring} have been widely adopted in consumer VR applications. Research has shown that controller-based text entry is often fast and easy to learn, with users reaching a mean typing speed of 15.4 WPM after only 5 minutes of practice~\cite{speicher2018selection}. While techniques with controllers can work well for VR applications that support a defined activity in a fixed space, the constant need to hold controllers and their limited battery life prevents widespread use in more ubiquitous AR situations.

\subsubsection{Eye Gaze}

Modern consumer HMDs (e.g., Hololens 2 and Meta Quest Pro) that support eye tracking presents subtle input using gaze-based techniques~\cite{lu2021itext, kurauchi2016eyeswipe, feng2021hgaze, cui2023glancewriter}. EyeSwipe~\cite{kurauchi2016eyeswipe}, for example, enabled gaze-only gesture-based typing with a mean typing speed of 11.7 WPM and a mean error rate of \textasciitilde1\% after 30 minutes of practice. Recently, Lu et al.~\cite{lu2021itext} proposed using head rotation for pointing and eye blinking for selection, achieving a mean typing speed of 13.8 WPM and a mean error rate of \textasciitilde1.5\% after 72 phrases of practice over four days. Although text entry with eye and head movements supports subtle interaction, such techniques impact natural viewing and suffer from eye fatigue~\cite{chitty2013user, kyto2018pinpointing, jalaliniya2014head,bernardos2016comparison}.

\subsubsection{Speech}

Text entry using speech recognition via the built-in microphones within HMDs has been considered an efficient method for text input~\cite{bowman2002text, ruan2018comparing, sawhney2000nomadic, price2004data}. Ruan et al.~\cite{ruan2018comparing} showed that a deep learning-based speech recognition system could achieve a mean text entry speed of 179 WPM and a mean error rate of \textasciitilde4.4\%. However, speech recognition is known to be unstable in noisy environments~\cite{price2004data}, and the use of speech raises privacy and social acceptability concerns~\cite{sawhney2000nomadic}.

\subsubsection{Mid-Air Hand Movement}

Several mid-air hand typing techniques have been proposed for unrestrained bare-hand AR activities. For example, Vulture~\cite{markussen2014vulture} enabled mid-air gesture swipe typing using a finger pinch that was tracked with an Optitrack motion tracking system. Although only word-level input was supported with this technique (i.e., letter-by-letter typing was not supported), it yielded a mean typing speed of 28.1 WPM with a mean error rate of \textasciitilde2\% after practice with 48 phrases. ATK~\cite{yi2015atk} used a Leap Motion sensor to enable mid-air ten-finger typing with a probabilistic tap detection algorithm for each finger. Likewise, although only word-level typing was supported, ATK was found to have a mean typing speed of 29.2 WPM and a mean error rate of 0.4\% after practice with 45+ phrases. ThumbAir~\cite{gil2023thumbair} typing using in-air two-thumb movements has recently shown a mean typing speed of 13.7 WPM with a mean error rate of 1.2\% after practicing 140 words and 35 phrases. Dudley et al.~\cite{dudley2018fast} proposed VISAR, which used direct finger touch on a mid-air virtual keyboard for both letter- and word-level typing. VISAR was found to have a mean typing speed of 17.8 WPM with a mean error rate of \textasciitilde0.6\% after practice with 80 phrases. It is currently the default text entry method in Microsoft Hololens 2. While these results are impressive, typing in mid-air lacks the haptic feedback that is generated when one touches a physical surface. The lack of surface haptics has been shown to induce significant arm and hand fatigue~\cite{cheng2022comfortable, jang2017modeling, arora2017experimental, hincapie2014consumed}. In addition, mid-air gestures typically require users to keep their hands raised at eye level, resulting in arm weariness and potential social acceptability concerns.

\subsubsection{Finger Taps on Physical Surfaces}

There has been active research to enable text entry on any physical surfaces around us, such as tables. Researchers have proposed the use of wearable Inertial Measurement Unit (IMU) devices, such as a ring~\cite{gu2020qwertyring}, two wristbands~\cite{streli2022taptype}, or two five-finger-straps~\cite{zhang2022typeanywhere} to decode typing sequences during finger tap typing on an imaginary QWERTY keyboard. An example of such an approach is TypeAnywhere~\cite{zhang2022typeanywhere}, which achieved a mean typing speed of 70.6 WPM with a mean error rate of 1.5\% after four days of practice. Although these outcomes are remarkable, these methods are mainly suited for word-level input due to their reliance on probablistic decoding at that level. For example, TapType \cite{streli2022taptype} evaluated typing performance using test phrases that include Out-Of-Vocabulary (OOV) words by treating each character as a ``word'', resulting in a considerable decrease in typing speed from 19.2 to 9.0 WPM.

\subsubsection{On-Skin Finger Touch}

Researchers have suggested using the skin as an always-available typing surface. Typing on the skin with bare-hand techniques also allows users to utilize their dexterous hand skills in an AR interaction. PalmType used the palm and fingers as input space and mapped each key of the QWERTY layout on separate segments of the skin~\cite{wang2015palmtype}. While eyes-free typing with an index finger, PalmType achieved a mean typing speed of 7.7 WPM. DigiTouch used the skin on one's finger for bimanual thumb-to-finger touch interaction (i.e., the thumb touched the skin of the fingers on the same hand), achieving a mean typing speed of 16.0 WPM and a mean error rate of \textasciitilde1\% after 180 minutes of practice~\cite{whitmire2017digitouch}. TipText~\cite{xu2019tiptext} and BiTipText~\cite{xu2020bitiptext} used conductive films attached at the first segment of the index finger to support eyes-free text entry (i.e., the thumb tip tapped on the film). Both techniques were limited to entering text only at the word level and did not offer the capability to input characters individually. The unimanual TipText achieved a mean typing speed of 13.3 WPM and a mean error rate of \textasciitilde0.3\%, whereas the bimanual BiTipText achieved a mean typing speed of 25 WPM and a mean error rate of \textasciitilde0.03\%. Both of these results were after practice with 30 phrases.

These techniques drastically differ from conventional typing methods, requiring users to learn new interaction paradigms often with custom hardware. STAR, on the other hand, offers a familiar typing experience to users by leveraging their physical smartphone two-thumb typing skills. Furthermore, STAR supports both word-level and letter-level input, offering practicality in real-life typing scenarios.

\section{STAR Design Process}

To design a smartphone-analogous typing method, we first surveyed how users typically hold and type on a physical smartphone. According to Palin et al.'s large-scale dataset on mobile text entry~\cite{palin2019people}, over 82\% of people used two thumbs to type. In addition, a public observation on how people hold mobile devices~\cite{steven2013How} found that people ``cradled'' their mobile phone in their fingers and used both thumbs for two-handed input (Figure \ref{fig:twoHandHoldPosture}). Within this context, we established the following \textbf{design principles}: 1) users should hold their hands in a similar manner to how they do while holding a mobile phone, 2) the same QWERTY layout should be used, 3) the finger movements used to type should be similar to smartphone typing, and 4) there should be a physical surface to type on as if typing on the smartphone.

Following these principles, during our design phase, we considered cognitive factors, including the intuitiveness of hand posture and physical factors, including the typing surface and keyboard layout and size.

\subsection{Elicitation Study: Typing Hand Posture}

\begin{figure}[t]
  \centering
  \includegraphics[width=8.5cm]{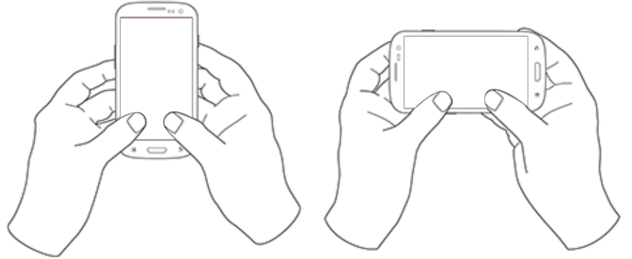}
  \caption{Hand postures during two-handed physical smartphone use that were observed during Hoober's studies (image from \cite{steven2022designing}, with permission).}
  \label{fig:twoHandHoldPosture}
\end{figure}

\begin{figure}[t]
  \centering
  \includegraphics[width=8.5cm]{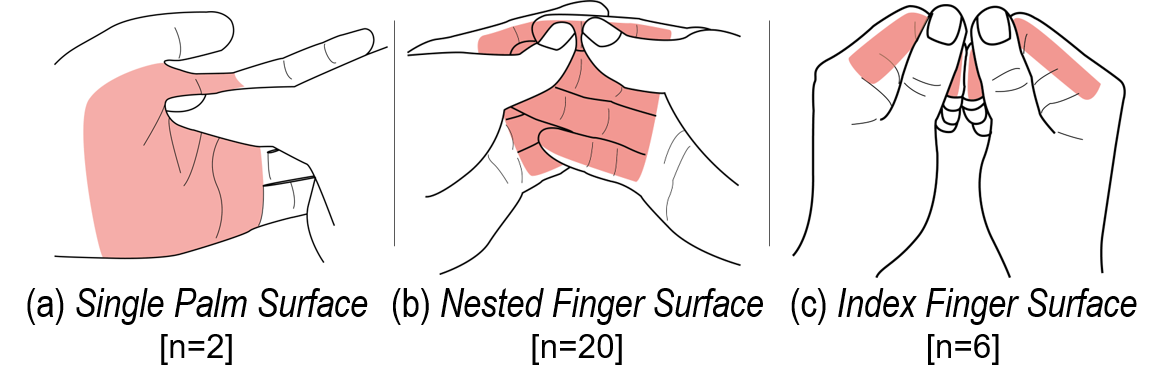}
  \caption{Results of the elicitation study depicting the skin surfaces and hand postures that users would use for imaginary smartphone typing. These are (a) Single Palm Surface with one hand and tapping with a finger from the other hand (n = 2), (b) Nested Finger Surface with both hands folded and tapping with both thumbs (n = 20) and (c) Index Finger Surface with the hands in a symmetric contact and tapping with both thumbs (n = 6). One participant used both thumbs while separating their hands, which did not fit into any category.}
  \label{fig:elicitationStudyResult}
\end{figure}

\begin{figure*}[t]
  \centering
  \includegraphics[width=0.78\textwidth]{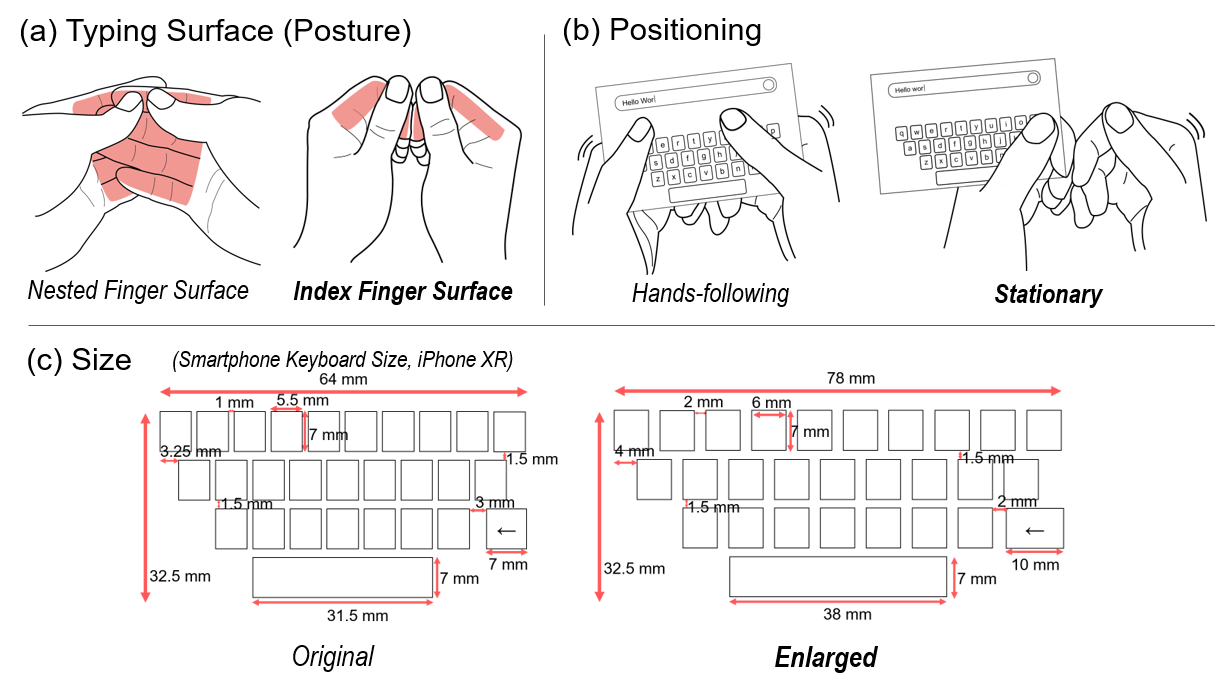}
  \caption{The three design parameters explored during the design phase for STAR: (a) the use of 
 hand surfaces for two-thumb typing such as the \textit{Index Finger Surface} and \textit{Nested Finger Surface}, (b) different Positionings such as \textit{Hands-Following} and \textit{Stationary}, and (c) different Sizes such as \textit{Original} (i.e., the iPhone XR's default keyboard layout) and \textit{Enlarged}. Boldfaced options were chosen for the final STAR design.}
  \label{fig:designOptions}
\end{figure*}

To explore the appropriate typing skin surface and corresponding hand posture for STAR, we conducted an elicitation study to observe the hand postures that users naturally used while typing on an imaginary smartphone. The elicitation study was run remotely through video calls with 29 participants (19 females, 10 males; age: M = 38 years, SD = 14 years). Participants were instructed to imagine that there was an invisible, imaginary smartphone in their hands to type on. Afterward, they were asked to type the sentence, \textit{"The quick brown fox jumps over the lazy dog"} on an imaginary smartphone in their hands. We observed hand postures from different angles. The study took about 15 minutes, including the introductory instructions and the completion of a demographic survey. All participants were paid. 

During the study, 93\% of participants (27 out of 29) used both thumbs for typing, and 96\% of them (26 out of 27) maintained contact between both hands while typing. This observation provided evidence that most users perceived typical smartphone typing as utilizing both thumbs while keeping the hands in contact. This is notable considering that users' hands may not always make contact during the use of a physical smartphone (Figure \ref{fig:twoHandHoldPosture}).

Many participants (n = 20) used the \textit{Nested Finger Surface} to create a thumb tapping space by placing their hands together in a folded position, with one hand on top of the other hand (Figure \ref{fig:elicitationStudyResult}b). The next most common approach (n = 6) was to utilize the \textit{Index Finger Surface} by making symmetric contact between the hands. In summary, the elicitation study revealed three main candidates for on-skin smartphone-analogous typing. As this research focuses on two-thumb typing, cases involving two thumbs (Figures 3b and 3c) were explored further.

\subsection{Prototype Development and Feedback}

During the iterative design of the technique, we focused on improving the accuracy, comfort, and efficiency by refining the typing surface, keyboard position, and keyboard size. 

\subsubsection{Typing Surface}

Based on the findings from the elicitation study, we singled out two candidates for on-skin two-thumb typing: the \textit{Nested Finger Surface} and the \textit{Index Finger Surface}. To examine the suitability of each typing surface in realizing STAR, we developed an initial prototype with a Hololens 2 AR HMD. The prototype utilized the HMD's vision-based hand tracking technology for thumb tracking and tap sensing. In the testing setup, a virtual keyboard of smartphone size (Figure \ref{fig:designOptions}c, \textit{Original}) was placed in the world space and displayed through the HMD. The participant then manually aligned their hand surface with the keyboard. The tracked position of the thumb's tip, which interacts with the keyboard, was visualized using a small colored sphere. 

During an informal pilot, we observed that users experienced difficulty in achieving stable key tapping on the virtual keyboard that is overlain on the  \textit{Nested Finger Surface}. This was due to the uneven typing surface caused by the depth difference between the two hands (i.e., one hand was placed behind the other). Moreover, the middle finger was often positioned slightly behind the index finger, adding uneven depth distance from each thumb. On the other hand, we observed that they were able to make more consistent key presses with better stability on the virtual keyboard that is overlain on the \textit{Index Finger Surface}. This could be attributed to the fact that the distance from the thumb to the surface was more even due to the flat property. Based on the pilot's feedback, we decided to use the \textit{Index Finger Surface} for on-skin thumb typing. Although the \textit{Index Finger Surface} was not the most popular choice from the elicitation study, we were encouraged that multiple participants (n = 6) independently suggested this posture as a canonical representation of imaginary smartphone typing, which also promoted more symmetric and ergonomic thumb-driven typing than the other option.

\subsubsection{Keyboard Positioning} \label{keyboardpositioning}

\begin{figure}[b]
  \centering
  \includegraphics[width=8.6cm]{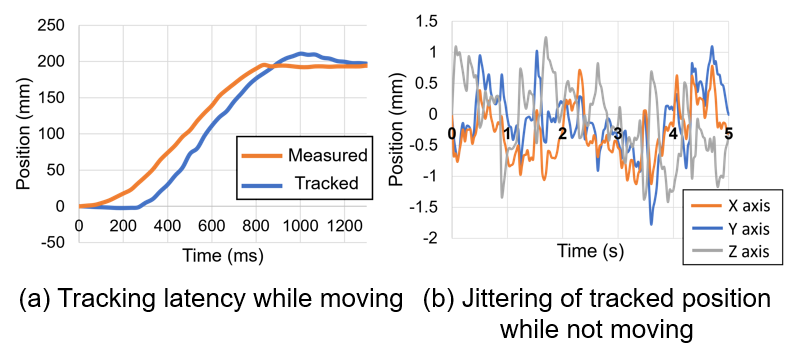}
  \caption{The observed (a) tracking latency and (b) tracking jitter of the thumb tip position from the Hololens 2 hand tracking. The tracking latency was measured while moving the hand approximately 200 mm along a single axis, whereas the measured position was approximated from the recorded video. (i.e., A ruler was put at a fixed position in the recorded video, and the tracked thumb was moved in line with the ruler.) The tracking jitter was measured for 5 seconds while the hand was held stationary on an armrest.}
  \label{fig:hololensTracking}
\end{figure}

When using a physical smartphone, the touchscreen follows a user's hand position as they are holding the device. We tried to mimic this experience by incorporating a \textit{Hands-Following} keyboard (Figure \ref{fig:designOptions}b), which updates its position and orientation according to the joints' tracked positions. To compensate for the hand tracking jitter (Figure \ref{fig:hololensTracking}b), a low-pass filter (i.e., 1-euro filter \cite{casiez20121}) was applied, and the axis of rotation was restricted for stable alignment between the keyboard plane and the user's \textit{Index Finger Surface}. During an informal pilot, however, we observed that users had difficulty tapping the intended keys as the hand tracking jitter was still causing the keyboard to shake. This was particularly noticeable when the two thumbs were continuously moving to type. Although a strong low-pass filter could alleviate this, it would also cause a delay in achieving timely synchronization between the virtual keyboard and the user's hand surface. We observed that even a tiny amount of unexpected displacement of the keyboard could significantly elevate users' tapping errors, when dealing with the small inter-key distances (1-2 mm). Therefore, we decided to try a \textit{Stationary} keyboard by fixing the keyboard at the initial position where the knuckle posture was made. With the \textit{Stationary} keyboard, a user can reposition the virtual keyboard by making a knuckle posture at a new location. 

During an informal test run, we observed that users were able to type more confidently using both thumbs on the \textit{Stationary} keyboard that is overlaid on the \textit{Index Finger Surface}, without worrying about the keyboard unexpectedly moving out of place. Finally, we decided to use the \textit{Stationary} positioning in our final implementation to simulate a stable keyboard alignment.

\subsubsection{Keyboard Size}

Lastly, we tested different keyboard layouts of varying sizes, starting with the layout of a physical smartphone keyboard (Figure \ref{fig:designOptions}c, \textit{Original}). During an informal assessment where users typed phrases they wanted on a \textit{Stationary} virtual keyboard overlain on the \textit{Index Finger Surface}, we observed that users frequently made tapping errors. This is anticipated, given that the keys were 5.5 mm wide with an inter-key distance of 1 mm, and the Hololens' finger tracking showed a tracking jitter of around $\pm$ 1 mm  (Figure \ref{fig:hololensTracking}b). We then designed larger keyboards that would facilitate more confident thumb tapping while making sure that keys on both ends (e.g., q or p) were within an easily reachable distance. Through an informal test, we examined various key widths (i.e., 6, 7, 8 mm) and inter-key distances (i.e., 2, 3, 4 mm), and decided to use a 6 mm key width and 2 mm inter-key distance for the \textit{Enlarged} keyboard size (Figure \ref{fig:designOptions}c) in our final implementation. Although we standardized the keyboard size for the controlled experiment, personalizing the keyboard size based on the user's hand size remains a valuable opportunity.

\section{STAR Implementation}

\begin{figure*}[t]
  \centering
  \includegraphics[width=1\textwidth]{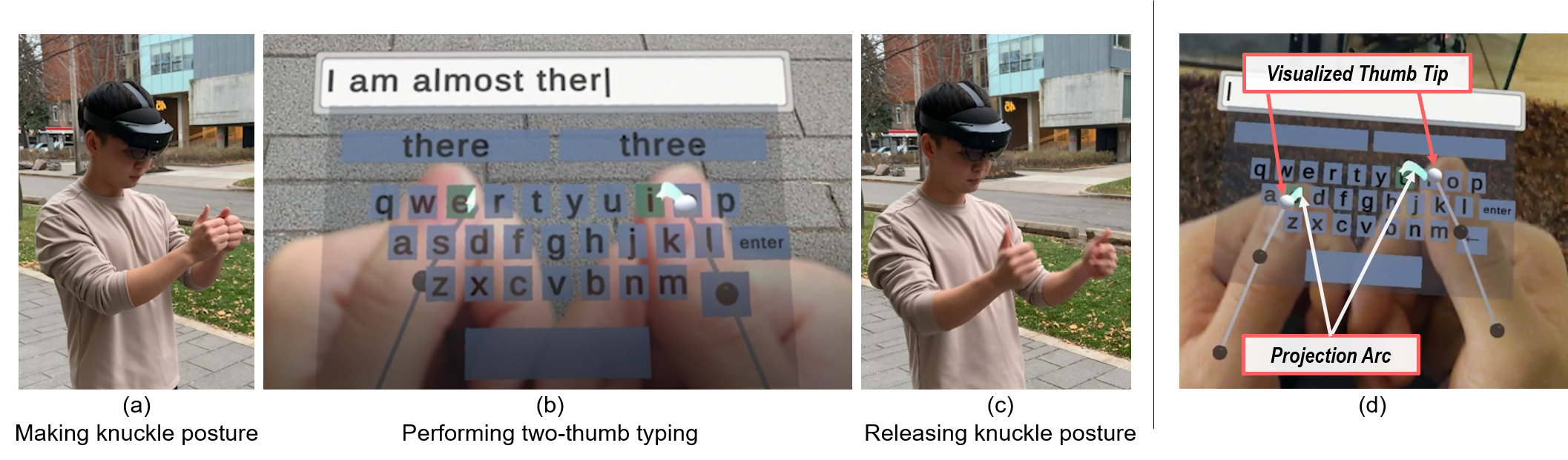}
  \caption{A typical interaction sequence while using STAR. (a) A user triggers STAR typing by making a knuckle posture. (b) The user performs familiar two-thumb typing on their \textit{Index Finger Surface} with visualized thumbs and projection arcs. (c) Finally, the user exits the typing mode by releasing the knuckle posture. (d) The clear view of the projection arc and visualized thumb joints.}
  \label{fig:STARInteraction}
\end{figure*}

Based on the explored design parameters, we arrived at a set of design decisions that optimizes usability and key input accuracy: \textit{Enlarged} size,  \textit{Index Finger Surface} typing surface, and \textit{Stationary} positioning as highlighted in bold texts in Figure \ref{fig:designOptions}. In this section, we describe the main components of our implementation: 1) hand tracking, 2) thumb tap sensing for key registration, 3) visualization details, and 4) word suggestion feature.

In the final implementation (Figure \ref{fig:STARInteraction}), a user can invoke STAR by making a ``knuckle posture'' as if holding a smartphone. The knuckle posture invokes the virtual keyboard and overlays it on the hands. The user can then perform two-thumb typing on the sides of the index fingers. Once the text entry is complete, the user can exit the typing mode by releasing the knuckle posture to seamlessly transition to other tasks. The knuckle posture is expected to support quick mode switching while preventing unintended activation since the sensory feedback of the hand contact signals the mode change~\cite{sellen1992prevention}.

\subsection{Bare-Hand Position Tracking}

We implemented STAR under the hand tracking capabilities of the current state-of-the-art AR HMD (i.e., Hololens 2). As modern AR/VR HMDs on the consumer market (e.g., Hololens 2, Meta Quest 2, and HTC Vive) support hand tracking for bare-hand interaction, the tracking fidelity is expected to be more mature over time. However, it currently has a noticeable tracking latency of \textasciitilde 90 ms and a tracking jitter with a range of approximately $\pm$1 mm, as shown in Figure \ref{fig:hololensTracking}. The selection of design parameters, such as \textit{Stationary} positioning, was made to simulate more stable hand tracking under the capabilities of the current state-of-the-art AR HMD (i.e., Hololens 2). Thus, there may be a chance to revisit the design decisions with a more reliable tracking environment.

\subsection{Thumb Tap Sensing (Key Registration)}

\begin{figure*}[t]
  \centering
  \includegraphics[width=0.69\textwidth]{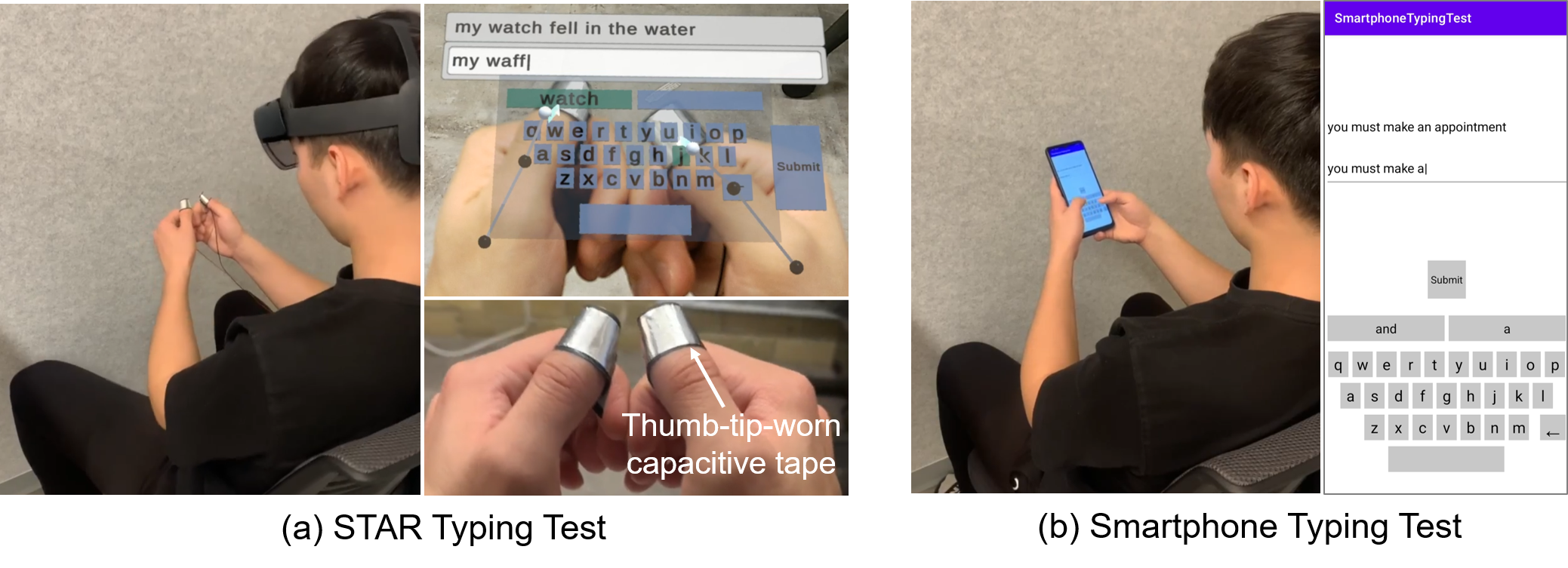}
  \includegraphics[width=0.30\textwidth]{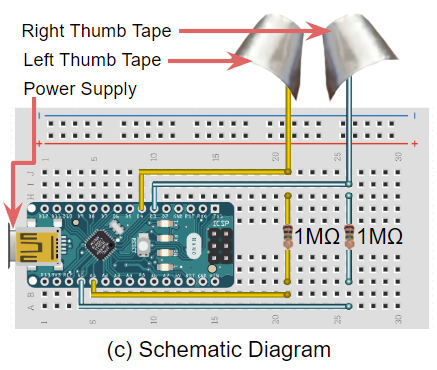}
  \caption{The testing environments used during the evaluation study. (a) In the STAR typing test, the participant put on the AR HMD (Hololens 2) and the thumb-tip-worn capacitive tapes. (b) In the smartphone typing test, the participant held a physical smartphone to type. (c) Schematic diagram includes an Arduino board, two 1M$\Omega$ resistors, and capacitive tapes for thumbs.}
  \label{fig:experimentalSetup}
\end{figure*}

Just like on a physical smartphone, a key tap should be detected whenever the thumb touches the \textit{Index Finger Surface}. We first implemented the detection of key tap (i.e., the contact between a thumb and the \textit{Index Finger Surface}) solely with vision-based hand tracking, as shown in Figure~\ref{fig:STARInteraction}. However, because the Hololens 2's finger tracking did not have sub-millimeter level precision, the key registration and thumb-finger contact was often temporally misaligned. This misalignment of haptic feedback led to errors and user complaints. To address this in the study, we developed robust thumb tap sensing beyond what the Hololens 2 could sense. We used thumb-tip-worn capacitive tapes (Figure \ref{fig:experimentalSetup}a) to simulate accurate tap detection. Thumb taps register the key with the closest center on the virtual keyboard. Given the continued advancement in hand tracking technology, we anticipate that fully bare-hand thumb tap sensing will achieve comparable performance levels. Further discussion on the realization of bare-handed tracking is in Section \ref{realWorldDeployment}.

\subsection{Visualization Details}

In see-through (i.e., glasses-based) augmented reality, where virtual pixels are overlain on the physical environment, users often confuse the ``depth'' of a virtual object \cite{ping2020effects}. We also observed that users often misjudged the distance between their hands and the virtual keyboard, which resulted in improper hand alignment. To alleviate the depth perception, we visualized the projection arc from the thumb tip to the keyboard plane, and highlighted the hovered key at the end of the arc using color feedback (Figure \ref{fig:STARInteraction}d). In addition, we visualized all three thumb joints and the connections between them above the keyboard, to provide users with a better perception of thumb depth. 

During an informal test, we found that users frequently made tapping mistakes when only the \textit{ThumbTip} joint was visualized. This was attributed to the thumb position shifting during curved tapping movements. For example, users first located their thumb straight above an intended key (e.g., k) and executed a thumb tap, resulting in them touching an adjacent key (e.g., j) due to the curved trajectory caused by the thumb's joint flexion (i.e., the human thumb tip does not move in a straight line, but rather an arc \cite{yi20232d}). After visualizing all three thumb joints and the expected projection arcs, we observed that users better predicted the key to be selected by their curved thumb tapping movements.

\subsection{Word Suggestion Feature}

The completion of a word via predictive suggestion is a commonplace feature within today's mobile text entry systems. There are three common types of suggestions: auto-completion/correction when the Space is pressed, suggestions that are tapped before one types all the letters in a word, and suggestions that are tapped to make a correction after typing all the letters in a word. Although widely used, forcing auto-completion/correction on every Space could interfere with the evaluation of the technique itself. Therefore, we included the second feature, which allowed participants to use completion by tapping suggested words based on their needs. The words with the highest and second highest probability were suggested on the left and right buttons, respectively.

To identify the two most probable words, a statistical decoding system combined spatial probability and language model probability. First, for each touch point on the keyboard, the spatial probability of each letter was calculated using a bivariate Gaussian Distribution \cite{azenkot2012touch} (e.g., t: 0.544, y: 0.432, ..., p: 0.001; the $\sigma$ value was the distance between the center of each key). By multiplying the spatial probabilities of each touch sample, the probabilities of the possible sequence of characters could be calculated (e.g., thw: 0.425, the: 0.312, thr: 0.266, ..., pmz: 0.000). The decoder then generated a list of words that begin with each candidate sequence of characters (as a prefix) in the language corpus, and multiplied the language model probability to calculate the final probability of the word. For the language corpus, Kaufman's lexicon \cite{Kaufman} was used, with each word's frequency extracted from Wikipedia corpora \cite{WikipediaCorpora}.

\section{STAR EVALUATION}

We conducted a user study to evaluate STAR's text entry performance. We also included a physical smartphone typing condition (\textit{Smartphone}) to observe the performance gap between \textit{STAR} and \textit{Smartphone}. As the goal of the study was not to outperform \textit{Smartphone} but to determine how close \textit{STAR} could approach state-of-the-art smartphone typing, \textit{Smartphone} was evaluated before and after the study was completed (i.e., it was not a typical within-subject factor). As the text entry speeds on the two instances of \textit{Smartphone} typing were not significantly different (39.9 vs. 39.0 WPM), they were averaged to form baseline data. 

\subsection{Participants}

Ten participants (i.e., two females, eight males; age: M = 25 years, SD = 4 years) were recruited from a university community to participate in the study. Two participants wore glasses, two participants were left-handed, and eight participants had no prior AR experience. All participants were paid.

\subsection{Apparatus}

A Microsoft Hololens 2, with a diagonal field of view of 52\textdegree{}, was used to run the Unity-based \textit{STAR} application. For thumb tap sensing, real-time capacitance data from the thumb-tip-worn tape was sent to a PC from an Arduino board through serial communication and was then sent to the Hololens 2 device through wireless TCP communication (Figure \ref{fig:experimentalSetup}c). We used distinct thresholding for debouncing the signal. The system identifies the tap engagement when the capacitance surpasses 250 (unit returned from CapacitiveSensor library \cite{capacitivesensor}), and it recognizes the completion of the tap when it falls below 200 units. For the \textit{Smartphone} baseline, a Galaxy A13 5G smartphone was used. The \textit{Smartphone} test application was developed in Android Studio and used the \textit{Original} keyboard layout (Figure \ref{fig:experimentalSetup}b).

\aptLtoX{\begin{table*}[t]
  \setlength{\tabcolsep}{4pt} 
  \begin{tabular}{cccccccc}
\toprule
    & Block 1 & Block 2 & Block 3 & Block 4 & Block 5 & Smartphone \\
    \midrule
    Text Entry Speed (WPM) & 17.4 (3.1) & 18.5 (2.6) & 19.9 (2.2) & 21.8 (2.8) & 21.9 (2.8) & 39.4 (7.4) \\
    UER (\%) & 0.1 (0.2) & 0.2 (0.3) & 0.3 (0.4) & 0.3 (0.7) & 0.3 (0.6) & 0.5 (0.7) \\
    CER (\%) & 7.5 (4.8) & 9.2 (6.9) & 9.0 (5.6) & 7.9 (6.2) & 8.8 (4.2) & 8.3 (3.4) \\
    IKI (ms) & 805 (203) & 697 (154) & 676 (175) & 617 (128) & 585 (100) & 315 (72) \\
    Backspace Usage (count) & 3.0 (2.0) & 2.9 (2.4) & 3.2 (2.3) & 2.9 (2.4) & 3.0 (1.4) & 2.6 (1.2) \\
    Key Press Duration (ms) & 148 (25) & 138 (22) & 140 (21) & 138 (20) & 132 (17) & 84 (9)   \\
    \bottomrule
    \end{tabular}
    \caption{The means and standard deviations of the text entry speeds, uncorrected error rates (UER), corrected error rates (CER), inter-key intervals (IKI), backspace usage counts, and key press durations during the user study. The performance of each block is reported separately for the \textit{STAR} condition, and the average of the two blocks is reported for the \textit{Smartphone} condition.}    
    \label{table:mainStudyResult}
\end{table*}}{\begin{table*}[t]
  \setlength{\tabcolsep}{4pt} 
  \begin{tabular}{cccccccc}
    & Block 1 & Block 2 & Block 3 & Block 4 & Block 5 & Smartphone \\
    \midrule
    Text Entry Speed (WPM) & 17.4 (3.1) & 18.5 (2.6) & 19.9 (2.2) & 21.8 (2.8) & 21.9 (2.8) & 39.4 (7.4) \\
    UER (\%) & 0.1 (0.2) & 0.2 (0.3) & 0.3 (0.4) & 0.3 (0.7) & 0.3 (0.6) & 0.5 (0.7) \\
    CER (\%) & 7.5 (4.8) & 9.2 (6.9) & 9.0 (5.6) & 7.9 (6.2) & 8.8 (4.2) & 8.3 (3.4) \\
    IKI (ms) & 805 (203) & 697 (154) & 676 (175) & 617 (128) & 585 (100) & 315 (72) \\
    Backspace Usage (count) & 3.0 (2.0) & 2.9 (2.4) & 3.2 (2.3) & 2.9 (2.4) & 3.0 (1.4) & 2.6 (1.2) \\
    Key Press Duration (ms) & 148 (25) & 138 (22) & 140 (21) & 138 (20) & 132 (17) & 84 (9)   \\
    \bottomrule
    \end{tabular}
    \caption{The means and standard deviations of the text entry speeds, uncorrected error rates (UER), corrected error rates (CER), inter-key intervals (IKI), backspace usage counts, and key press durations during the user study. The performance of each block is reported separately for the \textit{STAR} condition, and the average of the two blocks is reported for the \textit{Smartphone} condition.}    
    \label{table:mainStudyResult}
\end{table*}}

\begin{figure*}[t]
  \centering
  \includegraphics[width=1\textwidth]{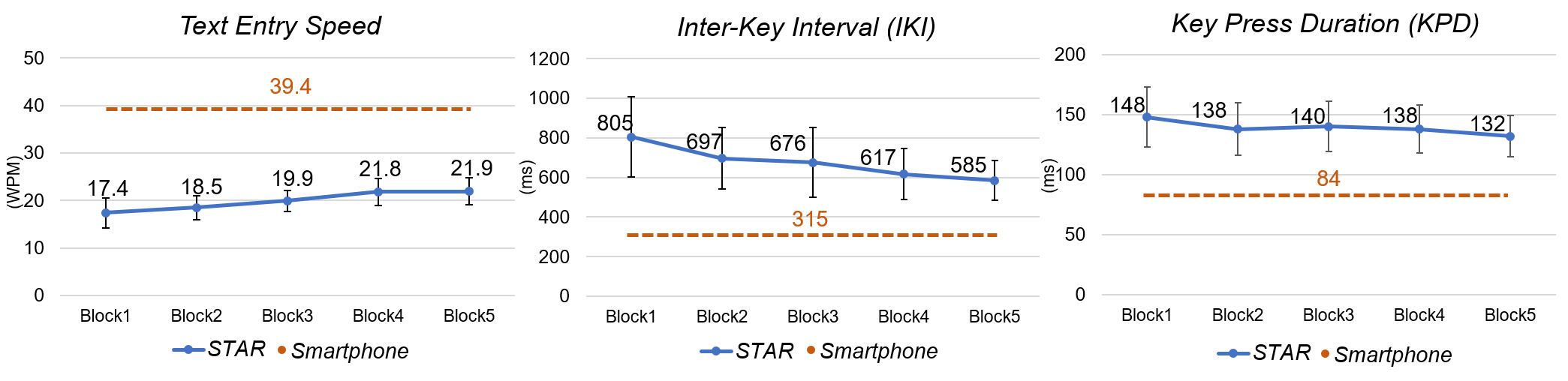}
  \caption{Line plots of mean Text Entry Speed, Inter-Key Interval (IKI)s, and Key Press Duration (KPD) from the evaluation (error bars show standard deviation).}
  \label{fig:resultPlot}
\end{figure*}

\subsection{Procedure}

At the beginning of the study, participants were asked to sit on a chair with no armrest. After completing a consent form, participants were instructed on how to complete the experiment using a video and slides. Participants were asked to roll up their sleeves so they would not cover their wrists and face a wall with a plain background to allow for stable hand tracking. Participants donned the thumb-tip-worn capacitive tape before starting the \textit{STAR} blocks. 

On each trial, participants were asked to transcribe a phrase that was randomly generated from the Mackenzie and Soukoreff phrase set \cite{mackenzie2003phrase}. Phrases that contained words that were not in Kaufman's lexicon \cite{Kaufman} were excluded. The set of phrases presented was identical for all participants. To transcribe a phrase, participants were instructed to make a ``knuckle posture'' to open the keyboard and then tap their ``\textit{Index Finger Surface}'' with a thumb to touch a key. Similar to many other evaluations of text input techniques \cite{dudley2018fast, palin2019people}, users could opt to complete a word by tapping on word suggestions, consisting of two options in our design (Figure \ref{fig:experimentalSetup}). After each transcription trial was complete, participants clicked on the \textit{Submit} button. They could then check their text entry speed and error rate for that trial. 

Each block in the study contained ten transcription trials. Participants completed one block of \textit{Smartphone} before completing the five \textit{STAR} blocks, and then a final block of \textit{Smartphone} after completing the \textit{STAR} blocks. Prior to the first \textit{STAR} block, participants were asked to transcribe three sample phrases to familiarize themselves with the \textit{STAR} application. At the beginning of each block, a preparation stage allowed participants to touch keys on an empty text field without knowing the target phrase to allow them to make minor adjustments to the height/distance of their hands to the keyboard if desired. Once a participant was ready to start transcribing, they pressed the \textit{Show Phrase} key (the \textit{Submit} button shown in Figure \ref{fig:experimentalSetup}a was \textit{Show Phrase} before starting a trial) to get the target phrase. Participants were informed that their typing speed would be recorded from the moment they pressed their first letter, so they could memorize the target phrase before starting the transcription if desired.

Participants were instructed to type as quickly and accurately as possible. Between each block, participants removed the HMD and had a break of at least two minutes. In total, the experiment took approximately two hours to complete.

\subsection{Metrics and Analysis}
Several metrics were computed to understand the degree to which STAR approached state-of-the-art smartphone typing.

Text entry speed and error rate were computed as they are the two main metrics of performance in text entry research. Text entry speed was measured in Words Per Minute (WPM), where the effective word count was calculated based on the number of transcribed characters minus one divided by a nominal word length of five \cite{wobbrock2007measures}. The entry duration was measured from the first key input to the last key input for each phrase. We computed three error rate metrics: the Uncorrected Error Rate (UER), the Corrected Error Rate (CER), and the number of times the backspace key was pressed \cite{soukoreff2003metrics}. The UER counted the errors in the submitted text based on the Minimum String Distance \cite{wobbrock2007measures}. The CER was similar to the UER but also counted backspace usage as an error.

Micro-metrics such as inter-key interval (IKI) \cite{feit2016we} and key press duration \cite{dudley2019performance} were also collected to obtain an in-depth analysis of typing behaviors. Inter-Key Interval (IKI) is the time between two subsequent key inputs \cite{feit2016we}, and can correlate with text entry speed. Key Press Duration (KPD) is the time between key down and key up events \cite{dudley2019performance}. For \textit{STAR}, the KPD was measured from the moment the thumb touched the \textit{Index Finger Surface} to the moment it was released. For \textit{Smartphone}, the KPD was measured from the firing of the key down event to the key up event as measured by the Android API.

To measure the learning effect over blocks, one-way RM ANOVAs were used for metrics with normal distributions (i.e., the text entry speed, IKI, and key press duration metrics) and Friedman tests were used for metrics without normal distributions (i.e., error rate and backspace usage count metrics). For post-hoc comparisons, paired sample t-tests with a Bonferroni correction were performed.

In addition to qualitative metrics, we collected participants' subjective feedback via the post-interview. In addition to their overall experience, we asked their thoughts on the main factors that affected their performance difference between \textit{STAR} and \textit{Smartphone}. 

\subsection{Results}
We first describe the quantitative results (Table \ref{table:mainStudyResult}) and then report on qualitative feedback from our study participants.

\subsubsection{Text Entry Speed}
The RM-ANOVA revealed significant differences across \textit{STAR} blocks (\textit{F}(4,36) = 10.996, \textit{p} < .001). Post-hoc comparisons revealed that Block 1 (17.4 WPM) was significantly slower than Block 4 (21.8 WPM; \textit{p} < .05) and 5 (21.9 WPM; \textit{p} < .001) and that Block 2 (18.5 WPM) was significantly slower than Block 5 (\textit{p} < .005). As the text entry speeds on the two blocks of \textit{Smartphone} typing were not significantly different (39.9 vs. 39.0 WPM), they were averaged to form a baseline (39.4 WPM). By Block 5, participants were performing at up to 56\% of the smartphone typing speed. Note that by the last block, the fastest typist reached 25.0 WPM with \textit{STAR}, whereas the slowest typist reached 16.4 WPM. 

\subsubsection{Error Rates}
The Friedman Test did not find any significant differences in UER across the five blocks of \textit{STAR}. The average UER was 0.2\%, whereas, for \textit{Smartphone}, the average UER was 0.5\%. These UERs indicated that participants were able to complete the transcription task with few errors during both conditions.

The Friedman Test did not find any significant differences in CER across the five blocks of \textit{STAR}. The average CER was 8.5\%, whereas, for \textit{Smartphone}, the average CER was 8.3\%. For backspace usage, the Friedman Test did not find any significant differences across the five blocks of \textit{STAR}. On average, participants used the backspace key 3 times during the \textit{STAR} condition, and 2.6 times during the \textit{Smartphone} condition. These similar CER and backspace usage results indicate that participants exhibited comparable degrees of correction behavior while using both \textit{STAR} and \textit{Smartphone} techniques.

\subsubsection{Inter-Key Interval (IKI)}
The RM-ANOVA found significant differences across blocks for IKI in \textit{STAR} (\textit{F}(4,36) = 12.703, \textit{p} < .001). As the mean IKI for \textit{Smartphone} was 315 milliseconds, by Block 5, participants' IKI was 585 milliseconds, which was 54\% of their smartphone performance and is consistent with the ratio of their text entry speed (i.e., 21.9 WPM / 39.4 WPM = 0.56).  

\subsubsection{Key Press Duration (KPD)}
The RM-ANOVA did not reveal significant differences across blocks for the time between key presses in \textit{STAR}, with the average KPD across all five blocks being 139 milliseconds. As the mean KPD for \textit{Smartphone} (84 milliseconds) was 1.6 times faster than the \textit{STAR}, participants made 1.6 times faster key presses on the \textit{Smartphone} touchscreen than \textit{STAR}. This observation on KPD is important in that it allows us to take a deeper look at the low-level factors influencing the performance gap. The discussion on this will be continued in Section \ref{performanceDelta}.

\subsubsection{Subjective Feedback}

In reference to their experience, some participants reported that they enjoyed using \textit{STAR}. P3, for example, mentioned \textit{"It was interesting that I am able to type without any on-hand device like a smartphone"} and P6 mentioned \textit{"It was overall a very new and pleasant experience"}. P7 also mentioned \textit{"It took some time to know how to type well, but it worked like smartphone typing once I got used to it."}

In response to the main causes of the performance issues with the techniques, five participants mentioned the limited fidelity of hand tracking. P6 noted that \textit{"the [visualized] thumb couldn't follow my speed when I moved my finger quickly from one key to another ... though it is a slight delay, it makes me keep checking the [visualized] thumb position to avoid mistakes"}. P5 and P6 commented on accidentally touched top-row keys (i.e., q, w, ..., p) while trying to touch the word suggestion button above them. P5 mentioned \textit{"When I quickly tap the word suggestion button, my (tracked) thumb wasn't following enough so the keys on the first row were selected instead"}.

Three participants wanted to use a version where the keyboard followed their hands and automatically aligned. Interestingly, this concept of the \textit{Hands-Following} keyboard was explored during our initial design process but was abandoned due to limited hand tracking fidelity (Section \ref{keyboardpositioning}). P7 mentioned \textit{"I had to think not only about typing but also about where my hand surface is relative to the keyboard. I think it slows down the typing"}. P7 also suggested using one-handed thumb swipe gesture typing, as many people do on smartphones today. Lastly, P3 suggested using mid-air gestures to trigger key input, e.g., performing a mid-air thumb left swipe for backspace.

\subsection{STAR vs. Smartphone: Performance Delta}
\label{performanceDelta}

STAR was able to achieve text entry speeds that were as high as 56\% of the users' physical smartphone typing speed, while maintaining similar error rates in terms of UER and CER with 30 minutes of training. As a text entry technique that leverages the same two-thumb typing skill as a physical smartphone, we believe that it has the potential to achieve even closer performance. Here, we analyze the factors contributing to the performance delta between STAR and smartphone typing using both quantitative performance metrics and subjective feedback.

\subsubsection{Hand Tracking}

The most crucial factor contributing to the performance gap was the HMD's unreliable hand tracking. As identified by half of the participants, it was necessary for users to keep checking the visualized thumb position to avoid mistakes. This can be confirmed again with quantitative metrics. First, the reported IKIs indicate that users on average spent 1.9 times longer (585 ms) key-to-key input than that of \textit{Smartphone} (315 ms). Users needed to keep monitoring the tracked thumb position to avoid errors, which eventually slowed down their key-to-key movement. Second, the reported KPDs indicate that users on average spent 1.7 times longer (139 ms) unit key press than that of \textit{Smartphone} (84 ms). Although it is presumed that the duration of a general thumb tap would not be much different either on an \textit{Index Finger Surface} or on a touchscreen, STAR showed considerably longer KPDs. We speculate that participants needed to wait for the lagging thumb marker to visually verify that the correct key had been pressed.

\subsubsection{Virtual Keyboard Positioning}

Another factor influencing the \textit{STAR} performance is the \textit{Stationary} positioning, which was dissimilar to how users employ smartphones. As outlined in Section \ref{keyboardpositioning}, the smartphone touchscreen follows the user’s hand position as users are holding it. Users can therefore leverage proprioceptive muscle memory when performing two-thumb typing, as each key in the keyboard is always at a fixed position relative to the hands. As we were unable to replicate this experience via a \textit{Hands-Following} keyboard due to the HMD's unreliable hand tracking, participants had to instead align their hands to the \textit{Stationary} keyboard while typing. P7 expressed concerns about having to pay constant attention to the hand position with respect to the keyboard, thus impacting the typing performance.

We again highlight that the limited hand tracking precision of the current state-of-the-art AR HMD was the most significant factor that restricted the performance of STAR. We believe STAR can reach typing performance closer to the smartphone over time with the continued advancement in HMD tracking technology. Incorporating additional sensing hardware that does not impede users’ thumb movement (e.g., rings \cite{chan2015cyclopsring, liang2023drg, zhang2017fingersound, ashbrook2011nenya}, smartwatches \cite{loclair2010pinchwatch}, or bracelets \cite{saponas2009enabling, dementyev2014wristflex}) may also open an opportunity to realize robust hand and finger tracking for STAR.

\section{Limitations and Future Work}
Our investigation into skill transfer for typing from smartphone to AR leads to several acknowledged limitations on our experimental setup due to the sensing technology, as well as opportunities for further investigation of the design space. 

\subsection{Real-World Deployment}
\label{realWorldDeployment}

To make STAR usable in real-world applications, the first step will be to achieve robust hand tracking and thumb tap sensing. This may prove difficult using only a HMD since parts of the hand can be occluded from the HMD's perspective. To address this issue, it would be beneficial to incorporate sensing hardware that does not impede the user's thumb typing movements (e.g., rings \cite{chan2015cyclopsring, liang2023drg, zhang2017fingersound, ashbrook2011nenya}, smartwatches \cite{loclair2010pinchwatch}, or bracelets \cite{saponas2009enabling, dementyev2014wristflex}). For example, wearing a technology like Electroring \cite{kienzle2021electroring} could immediately solve the challenge of thumb contact detection, although it would require users to wear rings on both hands. Further investigation of robust sensing techniques using cameras \cite{chan2015cyclopsring, loclair2010pinchwatch}, IMUs \cite{liang2023drg}, magnetic fields \cite{chen2013utrack, ashbrook2011nenya}, EMG \cite{saponas2009enabling}, sound \cite{zhang2017fingersound}, RF signals \cite{kim2021atatouch}, and pressure \cite{dementyev2014wristflex} may lead to more robust hand and finger tracking for bare-hand text input methods like STAR.

Another crucial aspect to consider is to understand STAR \textit{in-the-wild}. Although STAR was evaluated in a controlled lab study, text entry can occur in diverse situations, such as with primary tasks (e.g., listening to music or during a conversation), with various body poses (e.g., standing, sitting, leaning, or lying), with different activities (e.g., walking, or resting on a desk), or with dynamic real-world background. By evaluating STAR in a more extensive range of environments, we will discover additional opportunities and guidance for further refinement of the technique.

\subsection{Practical Standalone Prototype vs. Simulating the Limits of the Idea}

During our initial prototyping, we faced the trade-off between a practical standalone implementation that would show the technique's performance with today's standalone hardware and a proof-of-experience level prototype, simulating the limits of the technique with future technologies. While each approach has its respective advantages, we elected a balanced approach. The position sensing is enabled with standalone HMD hardware, while the contact detection is enhanced with the capacitive tapes. We believe this approach provides slightly higher external validity to our results than if we had used high-precision motion tracking equipment (e.g., Optitrack) for position sensing. In particular, several unobtrusive techniques could be utilized to achieve reliable contact sensing \cite{kienzle2021electroring, loclair2010pinchwatch}, whereas replicating motion tracking equipment accuracy is still unachievable with onboard HMD hardware.

However, it is important to note that the performance limits of the STAR method under an ideal technology (e.g., simulation with a motion capture system) have yet to be investigated. Since this aspect was not explored within the scope of this study, it presents an intriguing question for future research endeavors.

\subsection{Typing Performance Modeling with Tracking Precision}

The hand tracking latency of Hololens 2 was approximately 90 ms, and a tracking jitter was around $\pm$ 1 mm. Within this range of tracking performance, the STAR method achieved 56\% of the physical smartphone typing speed. To further explore the influence of tracking latency and error on STAR's typing performance~\cite{pavlovych2009tradeoff}, future research could involve systematically manipulating these variables in an experiment and analyzing their specific impact on STAR's performance. This may reveal the level of tracking precision required to reproduce smartphone-level typing performance in practice.

\subsection{One-handed STAR}

While our research primarily focused on two-thumb smartphone typing, one-handed typing is also prevalent on smartphones \cite{palin2019people, steven2013How}. One-handed typing can involve either gesture swipe typing \cite{zhai2003shorthand, kristensson2004shark2} or character-level tap typing. As one-handed tap typing is presumably slower, gesture typing may be more suitable for the one-handed STAR (i.e., the one-handed gesture typing technique on a virtual keyboard overlain on the hand skin surface). Given that previous one-handed gesture typing techniques in AR/VR \cite{liang2023drg, kern2023text, chen2019exploring, yu2017tap, markussen2014vulture} have been mostly explored using an indirectly mapped cursor visualized on a large keyboard layout through a HMD, it could be an attractive direction for future research to investigate how to transfer smartphone gesture typing skills to on-skin, one-handed AR text entry.


\section{Conclusion}
This research presented a novel bare-hand text entry method that is analogous to physical smartphone two-thumb typing. Unlike earlier techniques that utilize new metaphors and movement patterns, the proposed STAR technique leverages familiar typing behaviors by transferring the same thumb-typing skills we use with physical smartphones to the AR context. The proposed technique was implemented with the current state-of-the-art AR HMD, and the evaluation study showed that it supported efficient text entry performance (i.e., 21.9 WPM), which was up to 56\% of participants' physical smartphone typing speeds. As the tracking technologies for HMDs continue their rapid advancement, typing with STAR may approach the level of  performance seen in smartphone typing. This progress will open up a promising opportunity for STAR to become the preferred method of ubiquitous AR text entry.


\begin{acks}
We thank Michelle Annett for the proofreading, Rosa Morale for assisting in the study, and Brenton Rayner for the technical support. We also extend our appreciation to the numerous individuals from  Reality Labs Research and 2022 fellow interns for their insightful discussions.
\end{acks}


\balance
\bibliographystyle{ACM-Reference-Format}
\bibliography{main}

\end{document}